\documentclass[11pt,a4paper]{article}

\usepackage[margin=1in]{geometry}
\usepackage{amsmath,amssymb}
\usepackage{graphicx}
\usepackage{booktabs}
\usepackage{caption}
\usepackage{siunitx}
\usepackage[hidelinks]{hyperref}
\usepackage{setspace}
\usepackage{microtype}
\usepackage{float}
\usepackage{doi}

\graphicspath{{paper/figures/}{figures/}}

\newcommand{\lhat}{\hat{\lambda}}
\newcommand{\lhatMLE}{\hat{\lambda}_{\mathrm{MLE}}}
\newcommand{\mhat}{\hat{\mu}}
\newcommand{\dN}{\widehat{\Delta N}}
\newcommand{\Qzero}{Q_{0}}
\newcommand{\Qlam}{Q_{\lambda}}
\newcommand{\ppm}{\ensuremath{\mathrm{ppm}}}
\title{%
  Covariance scaling and a dual-mode torque diagnostic \\
  in measurements of the gravitational constant%
}

\author{Vineet Kumar\\
\href{mailto:v.kumar@usask.ca}{\texttt{v.kumar@usask.ca}}}
\date{October 3, 2026}

\begin{document}

\maketitle

% ============================================================
\begin{abstract}
\noindent
Laboratory determinations of the Newtonian gravitational constant $G$
disagree by more than their quoted uncertainties allow. We examine the
sixteen determinations comprising the CODATA-2018 archive using a
proportional covariance-scale model that retains the full reported
covariance matrix. Because the 2026 NIST torsion-balance
redetermination shares apparatus lineage with the BIPM Mark~II balance,
our primary historical fit excludes the BIPM-14 determination, giving
$\lhat = 2.967$ for the remaining fifteen measurements. Evaluated
against this historical model under the declared uniform
covariance-scaling convention, the correlated four-configuration NIST
result is highly unlikely under the archive's reported, unscaled
covariance but is statistically unremarkable once the historically
inferred extra dispersion is applied
($\Qlam = 4.840$; finite-sample $F$ $p = 0.382$); a mean-only scaling
sensitivity leaves the vector discrepant. Retaining the complete
archive also gives a comparable result. We further examine four matched
free-minus-servo torque differences reported by NIST. They are
consistent with a single common absolute torque offset across the four
tested configurations,
$\dN = \qty{1.918 \pm 0.215}{\pico\newton\metre}$, and poorly
described by an equally simple fractional-scale alternative under the
stated independent-Type-A treatment, with a chi-square contrast of
$\Delta\chi^{2} = 11.041$ between the two torque models. The paper
provides a covariance audit of the historical archive, a
lineage-qualified consistency check, and an experimentally testable
diagnostic for dual-mode torsion balances, without proposing a revised
value of $G$.
\end{abstract}

% ============================================================
\section{Introduction}
\label{sec:intro}

Among the fundamental constants of physics, the Newtonian gravitational
constant $G$ is the least precisely determined. Two centuries since the
Cavendish experiment, independent laboratory determinations continue to
disagree by substantially more than their individual reported
uncertainties allow \cite{Rothleitner2017,Li2018}. Contemporary
experiments have employed torsion balances operated in dynamic
(time-of-swing) and static (angular-deflection) modes, beam balances,
angular-acceleration feedback, pendulum displacement interferometry, and
atom interferometry \cite{Rothleitner2017,Li2018}, but the spread among
the accepted results remains several times larger than the individual
uncertainty budgets would suggest. CODATA has addressed this
discordance by applying a uniform multiplicative uncertainty expansion in its
recommended adjustment of $G$ rather than treating the individually
reported uncertainties and covariances as an adequate description of
the data \cite{Tiesinga2021,Mohr2025}. CODATA uses a 3.9 expansion factor to reduce all normalized residuals to 2 or less, following a long-established criterion of its adjustments.

Two recent statistical analyses have approached the resulting dispersion
from a Bayesian standpoint. Merkatas \textit{et al.} modelled the
CODATA archive with random-effects and Dirichlet-process mixture
techniques to obtain a consensus value under unrecognised, or ``dark'',
systematic uncertainty \cite{Merkatas2019}. Rinaldi \textit{et al.}
subsequently developed a hierarchical Bayesian treatment of systematic
errors in the $G$ record \cite{Rinaldi2023}. Those papers estimate consensus values under dark-uncertainty models. The present work instead asks a question of covariance-scale adequacy: whether the reported covariance structure of the CODATA-2018 archive, including its published off-diagonal terms that link determinations sharing apparatus, laboratory, or calibration history, is adequately summarised by a single multiplicative parameter, and, given the historical scale, whether a later apparatus-related four-configuration measurement is consistent with the archive.

The later NIST result was published after the CODATA-2018 adjustment,
but it is not an independent prospective replication. The NIST torsion
balance descends from the BIPM Mark~II apparatus, whose historical
result appears in the archive as BIPM-14
\cite{Quinn2013,Quinn2014,Schlamminger2026}. Our primary historical fit
therefore excludes BIPM-14. This reduces direct apparatus overlap
between the historical fit and the NIST vector, while not eliminating
shared design features or possible common systematic effects.

This work addresses three linked questions. First, under a
proportional-scale covariance audit, is the reported covariance of the
historical archive adequate? Second, under the declared
lineage-qualified historical model and predictive statistic, is the
later NIST four-configuration vector statistically unusual? Third,
under the stated independent-Type-A approximation, which of two
one-parameter torque-scaling descriptions better fits the matched
free-minus-servo torque data?

We show that the nominal reported covariance is inadequate for the
sixteen-determination CODATA-2018 archive under a proportional-scale
audit. Under the stated BIPM-14-excluded historical fit and the
declared predictive statistic, the later NIST configuration vector is
not statistically unusual once the historically inferred extra
dispersion is included. Under the stated Type~A uncertainty treatment,
the four NIST torque differences are better described by a common
absolute offset than by a one-parameter fractional-scale alternative.

% ============================================================
\section{Data}
\label{sec:data}

\subsection{Historical archive}

The determinations used here are the sixteen entries of the CODATA-2018 archive, together with their reported covariance structure \cite{Tiesinga2021}. The 2022 CODATA adjustment retained the same sixteen $G$ input determinations and the same expansion factor, so we refer to this historical input set by its original CODATA-2018 designation \cite{Mohr2025}. Reported standard uncertainties are used with coverage factor $k=1$ so
that the extra-dispersion scale $\lambda$ is estimated explicitly.

\subsection{NIST configuration vector}

The later measurement under study is the four-configuration result
reported by Schlamminger \textit{et al.} \cite{Schlamminger2026}: a
copper test mass and a sapphire test mass, each measured with the
torsion balance operated in electrostatic-servo mode and in free
(Cavendish) deflection mode. These four values are not independent;
they share the same suspension fibre, source masses, vacuum system,
and calibration chain, and their reported covariance matrix reflects
this. They are treated throughout as a single, correlated
four-component measurement, never as four separate determinations of
$G$.

% ============================================================
\section{Formalism}
\label{sec:formalism}

\subsection{A proportional covariance-scale model for the historical archive}

Let $\mathbf{y} = (y_{1},\dots,y_{N})^{\mathsf{T}}$ denote the vector of
$N$ historical determinations of $G$, and let $\mathbf{S}$ be the
corresponding $N \times N$ covariance matrix reported for the archive,
including all off-diagonal terms. We adopt the model
\begin{equation}
  \mathbf{y} \mid \mu,\lambda
  \;\sim\;
  \mathcal{N}\!\left(\mu\,\mathbf{1},\ \lambda^{2}\mathbf{S}\right),
  \label{eq:model}
\end{equation}
where $\mathbf{1}$ is the $N$-vector of ones, $\mu$ is a common
mean, and $\lambda > 0$ is a dimensionless covariance-scale factor
applied uniformly to the reported covariance. The archive is a curated
historical record rather than a random sample from a defined
population and therefore, the parameters inferred below are properties of this
specific record. The generalised least-squares estimator of $\mu$ is
\begin{equation}
  \mhat
  \;=\;
  \left(\mathbf{1}^{\mathsf{T}}\mathbf{S}^{-1}\mathbf{1}\right)^{-1}
  \mathbf{1}^{\mathsf{T}}\mathbf{S}^{-1}\mathbf{y},
  \qquad
  \operatorname{Var}(\mhat)
  \;=\;
  \left(\mathbf{1}^{\mathsf{T}}\mathbf{S}^{-1}\mathbf{1}\right)^{-1}.
  \label{eq:gls}
\end{equation}
The covariance-weighted residual quadratic is
\begin{equation}
  Q
  \;=\;
  \left(\mathbf{y} - \mhat\,\mathbf{1}\right)^{\mathsf{T}}
  \mathbf{S}^{-1}
  \left(\mathbf{y} - \mhat\,\mathbf{1}\right),
  \label{eq:Q}
\end{equation}
and the profile maximum-likelihood scale is
\begin{equation}
  \lhatMLE \;=\; \sqrt{Q / N}.
  \label{eq:lhatmle}
\end{equation}
The divisor $N$ gives the maximum-likelihood estimate under this model;
the profile likelihood in $\lambda$ has its maximum at
$\lambda = \sqrt{Q/N}$ after profiling over $\mu$. This is distinct
from the classical Birge ratio
\cite{Birge1932},
\begin{equation}
  R_{\mathrm{B}} \;=\; \sqrt{Q / \nu},
  \qquad
  \nu = N - 1,
  \label{eq:birge}
\end{equation}
which divides by the residual degrees of freedom remaining after
estimation of $\mu$, and from the empirical multiplicative expansion
factor adopted by CODATA in its recommended adjustment, the latter
being a policy choice that brings all normalized residuals under two. These three quantities
answer different consistency questions and are not interchangeable;
the profile scale and the Birge ratio are reported side by side in
Table~\ref{tab:historical}.

\subsection{A predictive statistic for the later measurement}

Let $\mathbf{y}_{\mathrm{N}}$ denote the four-component NIST vector and
$\mathbf{S}_{\mathrm{N}}$ its reported covariance matrix. No published
numerical cross-covariance links the historical archive to the NIST
vector; we therefore set that cross-covariance to zero as an explicit
analysis assumption. The nominal predictive covariance under the
unscaled historical model is
\begin{equation}
  \mathbf{V}_{0}
  \;=\;
  \mathbf{S}_{\mathrm{N}}
  +
  \operatorname{Var}(\mhat)\,
  \mathbf{1}_{4}\mathbf{1}_{4}^{\mathsf{T}},
  \label{eq:V0}
\end{equation}
where the rank-one term propagates the statistical uncertainty of the
historical mean into the comparison. The nominal discrepancy is
\begin{equation}
  \Qzero
  \;=\;
  \left(\mathbf{y}_{\mathrm{N}} - \mhat\,\mathbf{1}_{4}\right)^{\mathsf{T}}
  \mathbf{V}_{0}^{-1}
  \left(\mathbf{y}_{\mathrm{N}} - \mhat\,\mathbf{1}_{4}\right),
  \label{eq:Q0}
\end{equation}
and the corresponding scaled discrepancy under the fitted proportional
scale is
\begin{equation}
  \Qlam \;=\; \Qzero \big/ \lhatMLE^{2}.
  \label{eq:Qlam}
\end{equation}
For the primary screening calculation we apply $\lhatMLE$ uniformly to
both terms of Equation~\eqref{eq:V0}: to $\mathbf{S}_{\mathrm{N}}$ and
to $\operatorname{Var}(\mhat)$. This is a declared modelling
convention, not an independently estimated property of the NIST
experiment. Section~\ref{sec:nist-check} provides an internal-dispersion
check that is consistent with this convention but not decisive.

The upper-tail probabilities reported in
Section~\ref{sec:nist-check} refer to the finite-sample $F$ reference
under the stated independent Gaussian common-scale model. This is the
declared primary screening convention.%
\footnote{If the sampling variability of $\lhatMLE$ is treated
explicitly, the ratio $\Qlam$ follows a scaled $F$ distribution with
$(4, 14)$ degrees of freedom under the stated independent Gaussian
common-scale assumptions; the plug-in $\chi^{2}_{4}$ reference is
retained as a diagnostic, and the qualitative conclusion is unchanged.}

\subsection{Torque models for the dual-mode apparatus}

For each matched configuration $i$, define the free-minus-servo torque
difference
\begin{equation}
  d_{i} \;=\; N_{\mathrm{free},i} - N_{\mathrm{servo},i},
  \label{eq:delta}
\end{equation}
with combined Type~A standard uncertainty
\begin{equation}
  u_{i}
  \;=\;
  \sqrt{u_{\mathrm{free},i}^{2} + u_{\mathrm{servo},i}^{2}},
  \qquad
  w_{i} \;=\; u_{i}^{-2}.
  \label{eq:weights}
\end{equation}
The common absolute-offset model is
\begin{equation}
  d_{i} \;=\; \Delta N + \varepsilon_{i},
  \qquad
  \varepsilon_{i} \sim \mathcal{N}(0, u_{i}^{2}),
  \label{eq:abs-model}
\end{equation}
with weighted least-squares estimate
\begin{equation}
  \dN \;=\; \frac{\sum_{i} w_{i} d_{i}}{\sum_{i} w_{i}},
  \qquad
  \chi^{2}_{\mathrm{abs}}
  \;=\;
  \sum_{i} w_{i}\!\left(d_{i} - \dN\right)^{2}.
  \label{eq:abs-est}
\end{equation}
The competing fractional-scale model is
\begin{equation}
  d_{i} \;=\; f\,N_{\mathrm{free},i} + \varepsilon_{i},
  \label{eq:frac-model}
\end{equation}
with
\begin{equation}
  \hat f
  \;=\;
  \frac{\sum_{i} w_{i} N_{\mathrm{free},i} d_{i}}
       {\sum_{i} w_{i} N_{\mathrm{free},i}^{2}},
  \qquad
  \chi^{2}_{\mathrm{frac}}
  \;=\;
  \sum_{i} w_{i}\!\left(d_{i} - \hat f\,N_{\mathrm{free},i}\right)^{2}.
  \label{eq:frac-est}
\end{equation}
Both descriptions fit one parameter to the same four points. The two
models are not nested and thus, Wilks' theorem does not apply. Their
difference
\begin{equation}
  \Delta\chi^{2}
  \;=\;
  \chi^{2}_{\mathrm{frac}} - \chi^{2}_{\mathrm{abs}}
  \label{eq:dchi}
\end{equation}
is therefore reported as a descriptive contrast in goodness of fit
rather than as a likelihood-ratio test. A constant torque residual
gives the approximate scaling
\begin{equation}
  \frac{\Delta G}{G} \;\approx\; \frac{\dN}{N_{\mathrm{grav}}},
  \label{eq:scaling}
\end{equation}
whereas a fractional calibration error predicts an approximately
constant $\Delta G/G$. Equation~\eqref{eq:scaling} supplies the
physical signature used in Section~\ref{sec:torque}.

% ============================================================
\section{Historical covariance audit}
\label{sec:audit}

Equations~\eqref{eq:model}--\eqref{eq:birge} are evaluated for two
pre-specified configurations of the archive: the complete set of
sixteen determinations and the fifteen-determination subset excluding
BIPM-14. The BIPM-14-excluded fit is the primary configuration. The complete archive is retained as
a robustness comparison, to show that the primary conclusions do not
depend sensitively on the exclusion choice.

\begin{table}[H]
  \centering
  \caption{%
    Consistency of the CODATA-2018 archive under the proportional
    covariance-scale model. The final column gives $\mhat$ in units of
    $10^{-11}\ \mathrm{m^{3}\,kg^{-1}\,s^{-2}}$.%
  }
  \label{tab:historical}
  \small
  \begin{tabular}{lrrrrrrr}
    \toprule
    Configuration & $N$ & $Q$ & $\nu$
      & $p\,(\chi^{2}_{\nu})$ & $\lhatMLE$ & $R_{\mathrm{B}}$
      & $\mhat$ \\
    \midrule
    Complete archive & 16 & 195.679 & 15 & $1.60 \times 10^{-33}$ & 3.497 & 3.612 & 6.674300 \\
    BIPM-14 excluded & 15 & 132.088 & 14 & $2.63 \times 10^{-21}$ & 2.967 & 3.072 & 6.674227 \\
    \bottomrule
  \end{tabular}
\end{table}

\begin{figure}[htb]
  \centering
  \includegraphics[width=0.90\textwidth]{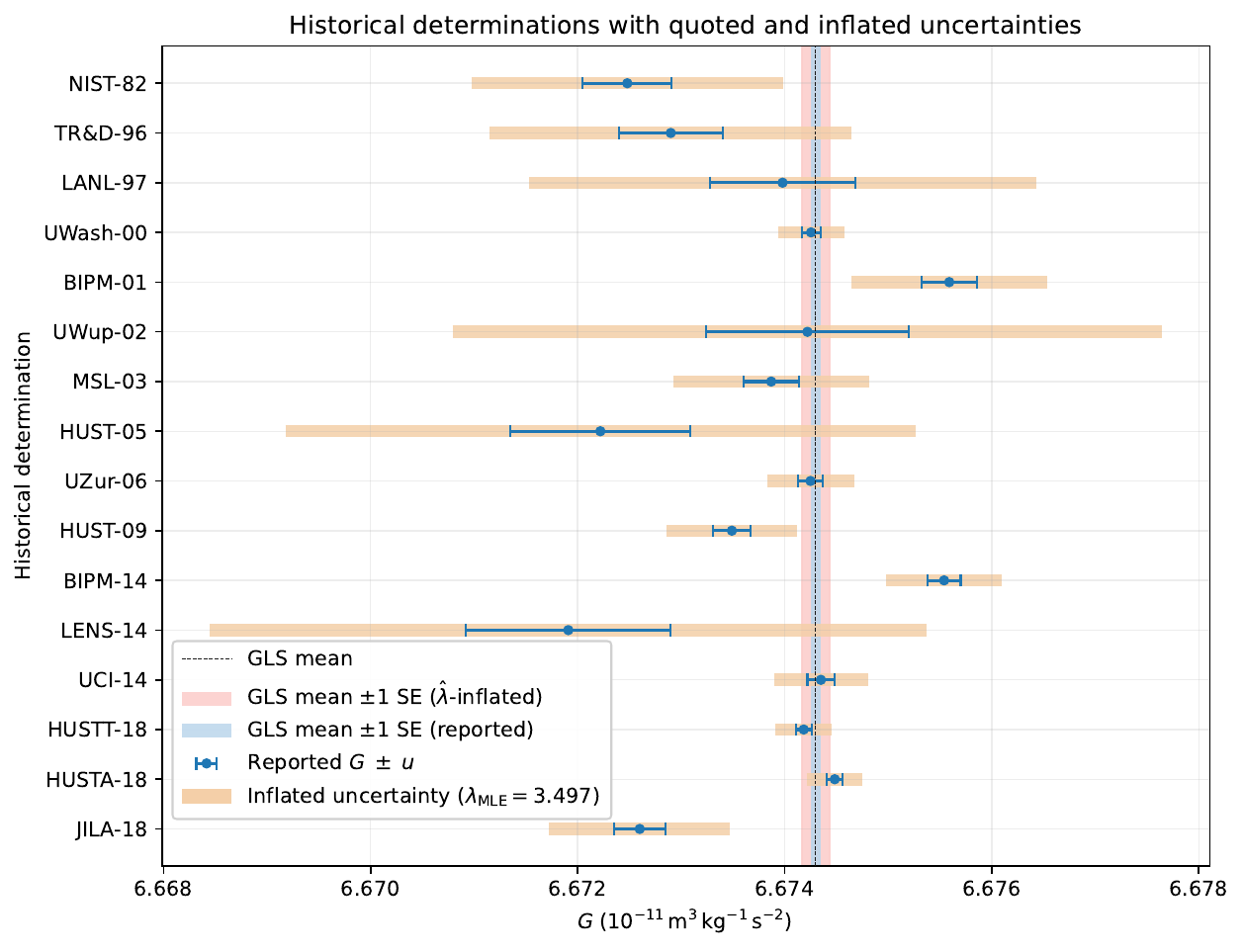}
  \caption{%
    Historical determinations of $G$ from the CODATA-2018 archive. For
    each determination, the narrow error bar is the quoted standard
    uncertainty and the wide error bar is the same uncertainty enlarged
    by the fitted covariance-scale factor $\lhat = 3.497$. The dashed
    vertical line marks the generalised least-squares mean for the
    complete archive.%
  }
  \label{fig:historical}
\end{figure}

The large inconsistency persists even after exclusion of BIPM-14: the
discrepancy statistic $Q$ remains overwhelmingly significant, and the
required covariance inflation remains close to a factor of three. Figure~\ref{fig:historical} shows the sixteen determinations with their quoted and inflated uncertainties.

To assess whether the proportional scale of Equation~\eqref{eq:model}
is a defensible choice among competing low-dimensional descriptions of
the excess dispersion, we compared it, on the complete 16-determination archive, with three
prespecified alternatives: an additive uncertainty floor, a
combination of proportional scaling and an additive floor, and a
power-law relation between the quoted and effective standard
uncertainty of each determination. The four models were compared by
the Bayesian information criterion (BIC), a standard penalised
likelihood criterion that balances goodness of fit against model
complexity.

For a model with maximised likelihood $\hat{L}$ and $k$ free
parameters fitted to $N$ observations, the criterion is
\begin{equation}
  \mathrm{BIC} = -2\ln\hat{L} + k\ln N ,
  \label{eq:bic}
\end{equation}
and the relative weights in Table~\ref{tab:bic} are proportional to
$\exp(-\Delta\mathrm{BIC}/2)$.

\begin{table}[H]
  \centering
  \caption{%
    BIC comparison of covariance-inflation models for the complete
    archive. Relative weights are normalized over the four models
    listed and are not model probabilities.%
  }
  \label{tab:bic}
  \small
  \begin{tabular}{lrrrr}
    \toprule
    Model & Free parameters & BIC & $\Delta$BIC & Relative weight \\
    \midrule
    Proportional scale   & 2 & 46.496 & 0.000 & 0.580 \\
    Power law            & 3 & 48.867 & 2.372 & 0.177 \\
    Proportional $+$ floor & 3 & 49.268 & 2.773 & 0.145 \\
    Additive floor       & 2 & 50.038 & 3.543 & 0.099 \\
    \bottomrule
  \end{tabular}
\end{table}

The proportional model is preferred within this restricted candidate
set. This establishes the proportional scale as a parsimonious working
description of the archive although it does not establish that the underlying laboratory errors are physically multiplicative and does not exclude other descriptions
outside the compared set. The proportional~$+$~floor model is not an ordinary interior fit: the fitted additive floor is effectively zero and lies on the boundary of
the admissible parameter space; therefore, the additional parameter is
not supported by this archive, and the BIC comparison for this model
carries the usual caveats associated with boundary estimates.

% ============================================================
\section{Consistency of the 2026 NIST result with the historical model}
\label{sec:nist-check}

The NIST measurement under study is the four-configuration vector
reported in Table~16 of Schlamminger \textit{et al.}~\cite{Schlamminger2026}. Table~\ref{tab:nist-values}
lists the four reported values and their relative standard
uncertainties; the analysis below uses the full reported $4\times 4$
covariance matrix (Table~18 of the same reference) rather than the
diagonal uncertainties alone.

\begin{table}[htbp]
  \centering
  \caption{%
    Reported NIST 2026 four-configuration $G$ values and their relative
    standard uncertainties ($k = 1$). The four values are correlated
    through shared apparatus and calibration; the full reported $4\times4$
    covariance matrix is used in the analysis.%
  }
  \label{tab:nist-values}
  \small
  \begin{tabular}{lrr}
    \toprule
    Configuration & $G\ (10^{-11}\ \mathrm{m^{3}\,kg^{-1}\,s^{-2}})$ & $u_{r}$ (\ppm) \\
    \midrule
    Copper, servo mode    & 6.673642 & 23.2 \\
    Copper, free mode     & 6.674021 & 30.3 \\
    Sapphire, servo mode  & 6.672637 & 37.5 \\
    Sapphire, free mode   & 6.673636 & 93.9 \\
    \bottomrule
  \end{tabular}
\end{table}

Equations~\eqref{eq:V0}--\eqref{eq:Qlam} are evaluated for this
four-component vector, giving the results in
Table~\ref{tab:prospective}.

\begin{table}[htbp]
  \centering
  \caption{%
    Prospective predictive comparison for the lineage-related 2026 NIST
    configuration vector. Uniform covariance scaling is the primary
    convention; mean-only scaling is a sensitivity analysis.
    Finite-sample $F$ tails are primary; plug-in chi-square tails are
    diagnostics. The quantity $\mhat$ is reported in units of
    $10^{-11}\ \mathrm{m^{3}\,kg^{-1}\,s^{-2}}$.%
  }
  \label{tab:prospective}
  \small
  \begin{tabular}{lrr}
    \toprule
    Quantity & BIPM-14 excluded (primary) & Complete archive \\
    \midrule
    Historical determinations                        & 15        & 16 \\
    $\mhat$                                          & 6.674227  & 6.674300 \\
    $\lhatMLE$                                       & 2.967     & 3.497 \\
    $\Qzero$ (reported covariance)                   & 42.625    & 47.172 \\
    $\Qlam$ (uniform scaling)                        & 4.840     & 3.857 \\
    Uniform tail, finite-sample $F$ (primary)        & 0.382     & 0.486 \\
    Uniform tail, $\chi^{2}_{4}$ plug-in (diagnostic)& 0.304     & 0.426 \\
    $Q$ (mean-only scaling)                          & 36.378    & 37.679 \\
    Mean-only tail, $\chi^{2}_{4}$ plug-in           & $2.4 \times 10^{-7}$ & $1.3 \times 10^{-7}$ \\
    \bottomrule
  \end{tabular}
\end{table}

Evaluated against the archive's reported, unscaled covariance, the NIST measurement
is highly unusual under the declared Gaussian screening convention but not when tested against the covariance scale inferred from the historical archive itself. The precise statement supported by this result is that,
under the stated exclusion of BIPM-14, the declared uniform
covariance-scaling convention, and the declared predictive statistic,
the four-configuration 2026 NIST result is not statistically unusual
once the historically inferred proportional extra dispersion is taken
into account. The finite-sample $F$ reference gives $p = 0.382$; the
plug-in $\chi^{2}_{4}$ reference gives $p = 0.304$ and is retained only
as a diagnostic. Under a mean-only scaling convention, in which the
historical scale is applied only to the historical-mean uncertainty,
the same NIST vector remains discrepant with the historical mean
($Q = 36.378$, plug-in $\chi^{2}_{4}$ $p = 2.4\times10^{-7}$).

The comparable result obtained when the complete archive is retained
(Table~\ref{tab:prospective}, right column) shows that this conclusion
does not depend sensitively on the BIPM-14 exclusion; the exclusion is
nonetheless retained as the primary configuration because it directly
addresses the apparatus lineage between the historical data and the
test measurement.

\begin{figure}[htb]
  \centering
  \includegraphics[width=0.90\textwidth]{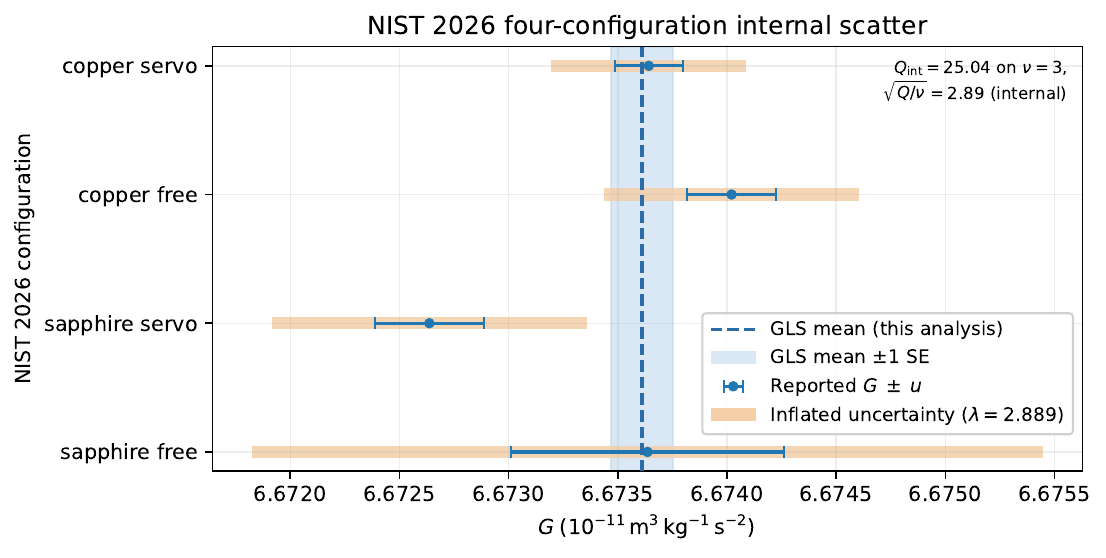}
  \caption{%
    Internal dispersion of the four reported 2026 NIST configuration
    values. The shaded band is the generalised least-squares mean of
    this analysis with $\pm 1$ standard error, computed from the
    reported $4\times 4$ covariance; it is not the published NIST
    Bayesian consensus. The inflated uncertainty bars enlarge each
    reported uncertainty by the NIST-internal dispersion factor
    $\sqrt{Q_{\mathrm{int}}/\nu} = 2.89$. The internal
    dispersion factor is separately computed from the same four values
    and is not used to define the predictive convention.%
  }
  \label{fig:nist-dispersion}
\end{figure}

As a supporting check, a generalised least-squares analysis of the
four reported NIST values against their reported covariance gives an
intercept-only residual quadratic
$Q_{\mathrm{internal}} = 25.04$ on $\nu = 3$ degrees of freedom,
corresponding to a Birge-like factor
$\sqrt{Q_{\mathrm{internal}}/\nu} = 2.89$. The internal NIST
dispersion factor $2.89$ is numerically close to the BIPM-14-excluded
historical scale $\lhatMLE = 2.967$, though the two quantify different
aspects of over-dispersion: the former is within a single
four-configuration apparatus, the latter is across the historical
archive. This is descriptive: it shows that the four NIST measurements are themselves over-dispersed relative to their reported covariance, at approximately the historical scale, but it does not by itself select the uniform-scaling convention declared in Section~\ref{sec:formalism}. Figure~\ref{fig:nist-dispersion}
displays the four reported values, the GLS mean
of this analysis, and the historically inflated uncertainty spans for
visual comparison.

% ============================================================
\section{Dual-mode torque diagnostic}
\label{sec:torque}

\subsection{Common absolute-offset model}

Table~15 of Ref.~\cite{Schlamminger2026} reports matched free- and
servo-mode torque measurements for three orientations of a copper
source mass and one sapphire configuration, giving four matched
configuration pairs in total. The corresponding differences and
combined Type~A uncertainties are given in Table~\ref{tab:torque}.

\begin{table}[htbp]
  \centering
  \caption{%
    Free-minus-servo torque differences reported for the four matched
    NIST configurations (Table~15 of Ref.~\cite{Schlamminger2026}).%
  }
  \label{tab:torque}
  \small
  \begin{tabular}{lrrr}
    \toprule
    Configuration & $d_{i}$ (\unit{\pico\newton\metre}) & $u_{i}$ (\unit{\pico\newton\metre}) & $N_{\mathrm{free},i}$ (\unit{\pico\newton\metre}) \\
    \midrule
    Cu, $0^{\circ}$   & 1.700 & 0.500 & \num{31197.9} \\
    Cu, $120^{\circ}$ & 1.400 & 0.671 & \num{31184.2} \\
    Cu, $240^{\circ}$ & 2.000 & 0.361 & \num{31185.6} \\
    Sapphire          & 2.100 & 0.361 & \num{13979.9} \\
    \bottomrule
  \end{tabular}
\end{table}

\begin{figure}[htb]
  \centering
  \includegraphics[width=0.90\textwidth]{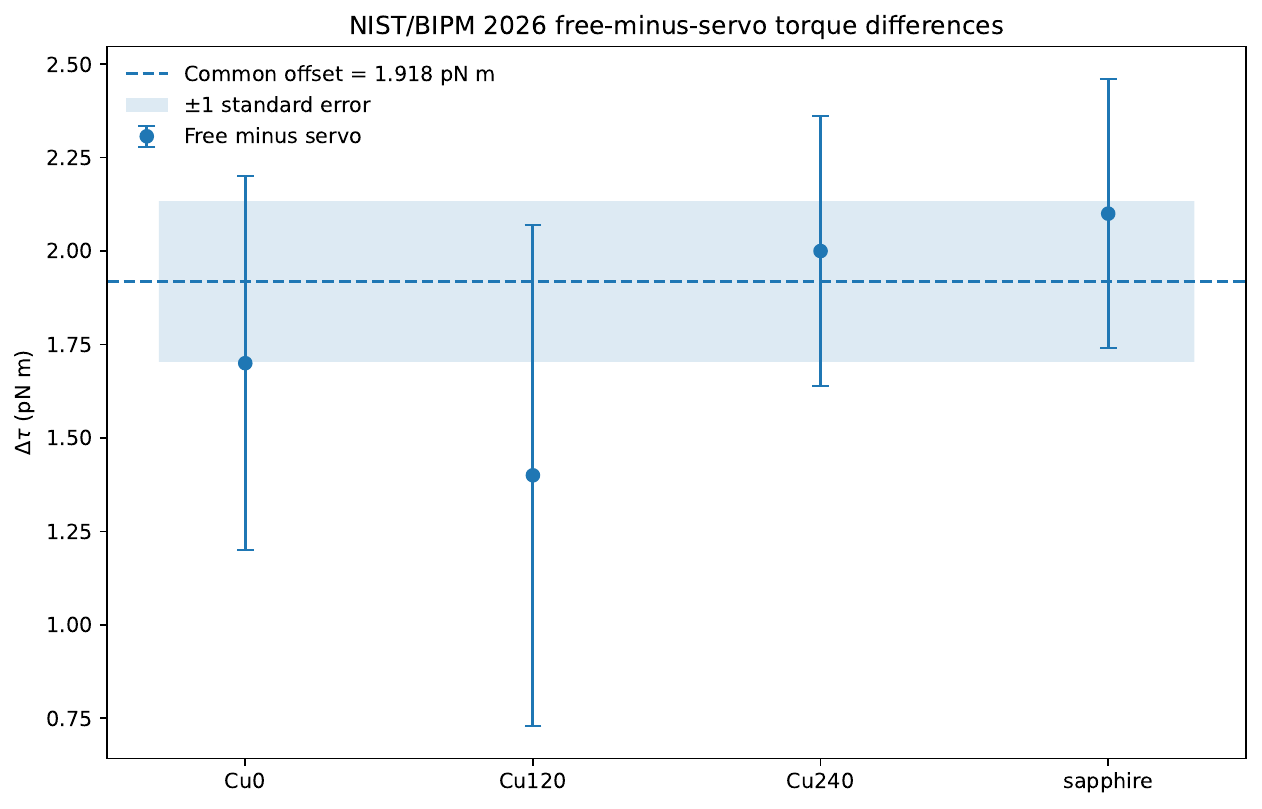}
  \caption{%
    Free-minus-servo torque differences for the four NIST
    configurations. The dashed horizontal line is the common
    absolute offset $\dN$; the shaded band is $\pm 1$ standard
    error.%
  }
  \label{fig:torque}
\end{figure}

Equations~\eqref{eq:abs-model}--\eqref{eq:abs-est} give a common
absolute offset of
\begin{equation}
  \dN \;=\; \qty{1.918 \pm 0.215}{\pico\newton\metre},
  \qquad
  \chi^{2}_{\mathrm{abs}} = 1.093\ (\nu = 3,\ p = 0.779).
  \label{eq:abs-result}
\end{equation}
The four torque differences are compatible with a common offset under
the independent-Type-A approximation (Figure~\ref{fig:torque}).
\footnote{The predictor $N_{\mathrm{free},i}$ is treated as fixed.
Because it is itself a measured quantity, treating it as error-free
slightly understates the standard error of the fitted fractional
coefficient $\hat f$, by a relative amount of order $10^{-4}$ at the
precision of the present data. This is negligible at the present
level of comparison and does not affect the reported contrast between
the two models. A symmetric reformulation in terms of the mean of
$N_{\mathrm{free},i}$ and $N_{\mathrm{servo},i}$ would change the
fitted model rather than correct this small errors-in-variables
effect, and is not pursued here.}

\subsection{Comparison with a fractional-scale alternative}

Equation~\eqref{eq:frac-model} applied to the same four points yields
\begin{equation}
  \hat f \;=\; 67.48 \pm 8.15\ \ppm,
  \qquad
  \chi^{2}_{\mathrm{frac}} = 12.134\ (\nu = 3,\ p = 0.0069).
  \label{eq:frac-result}
\end{equation}

\begin{table}[H]
  \centering
  \caption{%
    Comparison of the absolute-offset and fractional-scale torque
    models.%
  }
  \label{tab:torque-models}
  \small
  \begin{tabular}{lrrr}
    \toprule
    Model & Parameter & $\chi^{2}/\nu$ & $p$ \\
    \midrule
    Common absolute offset & $\dN = \qty{1.918 \pm 0.215}{\pico\newton\metre}$ & $1.093/3$ & 0.779 \\
    Fractional scale       & $\hat f = 67.48 \pm 8.15\ \ppm$ & $12.134/3$ & 0.0069 \\
    Difference             & $\Delta\chi^{2} = 11.041$ & --- & --- \\
    \bottomrule
  \end{tabular}
\end{table}

Because Equations~\eqref{eq:abs-model} and~\eqref{eq:frac-model} are
one-parameter models fitted to the same four measurements, the
difference $\Delta\chi^{2} = 11.041$ is reported as a descriptive
contrast in goodness of fit rather than as a formal likelihood-ratio
test. It nonetheless indicates a substantially lower chi-square for the
absolute-offset description under the stated independent-Type~A
approximation.

The physical basis for this preference follows from
Equation~\eqref{eq:scaling}. The sapphire configuration has a
free-mode signal torque of \qty{13979.9}{\pico\newton\metre}, roughly
45\% of the copper signal torque of \qty{31197.9}{\pico\newton\metre}.
Since the two source masses share the same geometry, the signal
ratio reflects the mass-density contrast: sapphire has a density of
approximately \qty{3.98}{\gram\per\cubic\centi\metre}, less than half 
of copper's \qty{8.96}{\gram\per\cubic\centi\metre}. A fixed absolute 
residual of \qty{1.918}{\pico\newton\metre} therefore corresponds to a 
fractional shift of approximately \qty{61}{\ppm} for copper and 
\qty{137}{\ppm} for sapphire, a ratio close to 2.2. This is comparable to the larger free-servo contrast reported for the sapphire configuration in the published values of $G$, where sapphire gives
$149.7\ \ppm$ and copper gives $56.8\ \ppm$, a ratio of approximately
2.6. The observed configuration
independence is consistent with an offset that does not depend on
source-mass material but it does not by itself exclude other mechanisms,
and the present data do not permit a mechanism identification.

As an internal-consistency check on the NIST four-configuration $G$
values, an additive material-plus-mode model compared with an
intercept-only model gives $\Delta\chi^{2} = 24.01$ on 2 degrees of
freedom ($p = 6.1\times10^{-6}$). The material-by-mode interaction is
not significant ($p = 0.311$). This supports the additive description
used above but is not an independent result.

The comparison has limits. The four differences share a common
apparatus and calibration chain; their true covariance is not, in
general, zero. A full covariance matrix for the four differences has
not been published, and the present treatment approximates the four
uncertainties as independent with $N_{\mathrm{free},i}$ held fixed.
The result is accordingly an apparatus-level descriptive diagnostic
rather than a complete propagated-uncertainty analysis.

The torque data constrain only the free-minus-servo difference
$\Delta N$, not how that difference is divided between a free-mode
bias and a servo-mode bias. Attributed entirely to the free mode, the
correction would shift the free-mode $G$ values downward by
$\Delta N/N_{\rm grav}$ and leave the servo values unchanged. The
correction would be approximately $-61$ ppm for copper and $-137$ ppm
for sapphire, the same fractional shifts derived above from the torque analysis.
Attributed entirely to the servo mode, the shifts apply to the servo
values with the opposite sign. Any intermediate division is equally
consistent with the torque data, and the present data do not
determine which applies. Their model includes a configuration-specific effect for each of the
four determinations \cite{Schlamminger2026}, whereas the model
considered here uses a single offset shared across all four. Both
descriptions account for the observed free-minus-servo contrast, and
the present data do not determine which parametrisation is correct.

\subsection{Candidate mechanisms and a discriminating experiment}

The NIST group reports a gas-mediated thermal torque of order
\qty{1.4}{\pico\newton\metre}, comparable in magnitude to $\dN$, but
observes no significant time dependence after applying the
servo voltage; this is an order-of-magnitude consistency check, not
evidence of causation. The reported capacitance and voltage uncertainties,
each of order \qty{6}{\ppm} in servo mode, do not provide a signed
residual capable of reproducing a bias of
\qty{1.9}{\pico\newton\metre}. Neither anelastic relaxation of the
suspension fibre nor patch-potential electrostatic effects can be
confirmed or excluded from the published record.
Table~\ref{tab:mechanisms} summarises the reported scales.

\begin{table}[htbp]
  \centering
  \caption{%
    Reported apparatus-level scales for candidate torque-error
    mechanisms. None supplies a signed residual capable of predicting
    the fitted \qty{1.918}{\pico\newton\metre} offset.%
  }
  \label{tab:mechanisms}
  \small
  \begin{tabular}{llc}
    \toprule
    Candidate mechanism              & Reported scale                     & Signed prediction? \\
    \midrule
    Gas-mediated thermal             & $\sim \qty{1.40}{\pico\newton\metre}$ & No \\
    Capacitance (servo mode)         & $\sim \qty{6}{\ppm}$               & No \\
    Voltage (servo mode)             & $\sim \qty{6}{\ppm}$               & No \\
    Autocollimator non-linearity     & Broad support interval             & No \\
    \bottomrule
  \end{tabular}
\end{table}

The sign of the observed offset is itself a constraint. Across the reported mode-level comparisons for the BIPM Mark~I, Mark~II, and NIST~2026 measurements, the free-deflection value exceeds the servo value, though the Mark~I contrast is consistent with zero at its own precision. Any candidate mechanism would therefore have to reproduce this sign as
well as the approximate magnitude.

Equation~\eqref{eq:scaling} suggests a direct experimental test: a
dual-mode balance in which the source-mass separation, and hence
$N_{\mathrm{grav}}$, is deliberately varied on a single apparatus
would produce $\Delta G/G \propto N_{\mathrm{grav}}^{-1}$ under the
absolute-offset hypothesis, and an approximately constant
$\Delta G/G$ under the fractional-scale hypothesis.
Table~\ref{tab:discriminator} gives the predicted fractional contrast
at three illustrative gravitational torque levels, with both
parameters fixed at their fitted values.

\begin{table}[htbp]
  \centering
  \caption{%
    Predicted fractional $G$ contrast under two hypotheses at three
    illustrative gravitational torque levels. The constant absolute
    offset is fixed at the fitted value
    $\dN = \qty{1.918}{\pico\newton\metre}$; the fractional scale is
    fixed at the fitted value $f = \qty{67.5}{\ppm}$.%
  }
  \label{tab:discriminator}
  \small
  \begin{tabular}{lrrr}
    \toprule
    Hypothesis
      & $N_{\mathrm{grav}} = \qty{14}{\nano\newton\metre}$
      & $\qty{31}{\nano\newton\metre}$
      & $\qty{62}{\nano\newton\metre}$ \\
    \midrule
    Constant absolute offset        & \qty{137}{\ppm} & \qty{62}{\ppm} & \qty{31}{\ppm} \\
    Fractional calibration error    & \qty{67.5}{\ppm} & \qty{67.5}{\ppm} & \qty{67.5}{\ppm} \\
    \bottomrule
  \end{tabular}
\end{table}

The four existing configurations already provide modest discriminating
power between the two scaling descriptions under the stated Type~A
approximation ($\Delta\chi^{2} = 11.041$). A purpose-built dual-mode
experiment in which the source-mass separation is deliberately varied
would sharpen this discrimination substantially.

% ============================================================
\section{The BIPM apparatus lineage}
\label{sec:bipm}

The NIST balance is the BIPM Mark~II apparatus, transferred to NIST
in 2016 and subsequently modified for the present measurement. Mark~II
was, in turn, a rebuild of the Mark~I apparatus. Both
earlier marks were operated in Cavendish free-deflection and
electrostatic-servo modes, and their reported mode contrasts provide
historical context for the NIST torque diagnostic.

The BIPM Mark~II balance reported final mode values
\cite{Quinn2013,Quinn2014}
\begin{equation}
  G_{\mathrm{Cav}}   = 6.67586(36)\times 10^{-11}\ \mathrm{m^{3}\,kg^{-1}\,s^{-2}},
  \qquad
  G_{\mathrm{servo}} = 6.67515(41)\times 10^{-11}\ \mathrm{m^{3}\,kg^{-1}\,s^{-2}},
  \label{eq:bipm}
\end{equation}
where the numerals in parentheses are uncertainties in the last two
mantissa digits, corresponding to relative uncertainties of $54$ and
$61\ \ppm$ respectively. With the reported mode correlation of
approximately $-0.63$, the covariance-aware contrast between the two
modes is $106 \pm 104\ \ppm$ (Cavendish higher), in the same sense as
the NIST result. Using the BIPM apparatus's stated peak-to-peak
gravitational signal-torque scale
($\sim \qty{31.5}{\nano\newton\metre}$) solely as a unit conversion,
this contrast corresponds to a torque-equivalent difference of
$\qty{3.3 \pm 3.3}{\pico\newton\metre}$.%
\footnote{Under the single-sided amplitude convention
($\sim \qty{17}{\nano\newton\metre}$), the same ppm contrast corresponds
to $\qty{1.8 \pm 1.8}{\pico\newton\metre}$. The peak-to-peak
convention is adopted here so that the comparison matches the
convention used by NIST~2026 when quoting its own BIPM comparison
value of $\qty{3.3}{\pico\newton\metre}$. The ppm contrast is
independent of the convention.}
The same experiment's combined two-mode result appears in the
CODATA-2018 archive as the BIPM-14 entry
$6.67554(16)\times 10^{-11}\ \mathrm{m^{3}\,kg^{-1}\,s^{-2}}$;
the two-mode values quoted above and this combined value refer to the
same apparatus and the same set of measurements.

For comparison, the earlier BIPM Mark~I apparatus
\cite{Quinn2001} reported mode values of
$6.67565(45)\times 10^{-11}$ for the Cavendish mode and
$6.67553(40)\times 10^{-11}$ for the servo mode, a covariance-aware
contrast of $18 \pm 98\ \ppm$ (Cavendish higher), consistent with zero
at its own precision. The Mark~I contrast is numerically
smaller than the Mark~II and NIST contrasts, although its uncertainty
does not formally exclude a similarly sized effect. The reported mode contrast is larger for the Mark~II measurements than for the Mark~I measurements, although the uncertainties are substantial. The present data do not identify whether any such difference is
associated with a specific change in the apparatus, and we do not pursue
a mechanism attribution.

This comparison is offered as a historical consistency check, not as
an independent replication. BIPM publications do not report
configuration-resolved raw torque residuals with their associated
Type~A uncertainties, and the BIPM mode contrasts remain statistically
compatible with zero at their own quoted precisions. The BIPM mode
comparison is consistent in sign and order of magnitude with the NIST
torque diagnostic, but it is neither configuration-resolved nor
statistically decisive and is not used as an independent test of the
NIST model.

% ============================================================
\section{Discussion}
\label{sec:discussion}

\subsection{Interpretation}

The inconsistency in the historical archive with its own quoted covariance requires an inflation factor near three and is not removed by discarding a single, lineage-implicated entry. Once this
inflation is applied consistently under the declared uniform-scaling
convention, the later, apparatus-related NIST measurement is no longer
an outlier. These two statements together constitute a
covariance-audit result: they characterise the statistical
structure of disagreement in the $G$ literature without asserting a
physical origin for that disagreement and without treating any
individual measurement as anomalous or in error.

\subsection{Relation to the CODATA adjustment}

This analysis reinterprets the CODATA-2018 archive, but does not produce
an alternative recommended value. CODATA's own uncertainty expansion
already reflects the inconsistency documented here. The scale factor
$\lhatMLE$ provides an explicit, likelihood-based quantification of
the same inconsistency for the specific archive and covariance model
considered. Neither this analysis nor the CODATA adjustment identifies
a physical cause for the excess dispersion.

Our results reproduce CODATA's own internal statistics once the expansion factor is removed. CODATA reports a post-expansion $\chi^{2} = 12.9$
on $\nu = 15$ for the sixteen-determination $G$ archive, with
$R_{\mathrm{B}} = 0.93$ after expansion. Multiplying the residual
quadratic by the square of the expansion factor gives
$12.9 \times 3.9^{2} \approx 196.2$, matching the unexpanded
$Q = 195.679$ reported in Table~\ref{tab:historical} to the precision
of the published values; correspondingly,
$R_{\mathrm{B}} \times 3.9 \approx 3.63$, matching our unexpanded
$R_{\mathrm{B}} = 3.612$. The profile-likelihood scale $\lhatMLE = 3.497$ obtained here by an
independent criterion is lower than but of comparable magnitude to
CODATA's $3.9$ expansion factor, which was chosen by a different
criterion: requiring all normalized residuals to be below two, a
threshold corresponding approximately to the 95\% central interval of a
standard Gaussian. The two estimates therefore are in agreement with each other, and this audit can be read as a likelihood-based reinterpretation of CODATA's own
adjustment.

\subsection{Scope of the numerical checks}

The Gaussian formalism of Section~\ref{sec:formalism} constitutes the
complete inferential basis for the results reported in
Sections~\ref{sec:audit}--\ref{sec:torque}. The primary narrative uses
profile-likelihood generalised least squares because it is
closed-form, metrologically transparent, and reproducible without a
stochastic sampler. A separate hierarchical Student-$t$ treatment of
the same archive was implemented as a robustness check and validated
by simulation, recovering the injected parameters of synthetic data
sets with the expected nominal coverage. This provides an independent
check on the associated inference code and on the sensitivity of the
conclusions to the Gaussian likelihood assumption, but it does not
modify any claim advanced in this paper.

The primary screening $p$-values reported in Section~\ref{sec:nist-check}
are finite-sample $F$ upper tails; the plug-in $\chi^{2}_{4}$ tails are
retained as diagnostics. Neither is a posterior-predictive tail.
Additional posterior-predictive calculations are retained in the
reproducibility package as supplementary checks and are not used as
primary evidence. These auxiliary calculations give the same
qualitative classification of the NIST vector under the alternative
hierarchical treatment, so the conclusion that the vector is not
unusual relative to the historical archive does not depend on that
supplementary model.

\subsection{Implications for future determinations}
\label{sec:forward}

The Gaussian model of Section~\ref{sec:formalism} also yields a
falsifiable prediction for future determinations of $G$. If the
proportional excess-dispersion scale inferred from the archive is also
applicable to future determinations, then a future independent
determination reporting value $y_{\rm fut}$ with quoted standard
uncertainty $u$ at $k = 1$ should have a typical deviation from the
historical mean of order
\begin{equation}
  |y_{\rm fut} - \mhat| \;\approx\; \lhatMLE\, u ,
  \label{eq:falsifiable}
\end{equation}
where $\mhat$ and $\lhatMLE$ are the historical GLS mean and the
fitted proportional scale of Section~\ref{sec:audit}.\footnote{More formally, under the same Gaussian common-scale model, the conditional 95\% predictive interval is $y_{\rm fut}=\mhat\pm c_F\,\lhatMLE\sqrt{u^2+\operatorname{Var}_{\rm rep}(\mhat)}$, where $c_F=\sqrt{(N/\nu)F_{0.95}(1,\nu)}$ and $\nu=N-1$. The prediction is conditional on the future determination's quoted $k=1$ uncertainty $u$. The finite-sample calibration is exact under the declared common-scale model, including the assumption that the same multiplicative scale applies to the future quoted uncertainty.}
Agreement at the level of the future determination's own quoted
uncertainty, $|y_{\rm fut}-\mhat| \approx u$, would therefore provide a
direct test of whether the historical proportional under-reporting
extrapolates to an independent future determination.

\subsection{Limitations}

The exclusion of BIPM-14 mitigates, but does not eliminate, the shared
apparatus lineage between the historical archive and the
later NIST measurement and thus, a genuinely independent redetermination,
obtained with an unrelated experimental technique, remains the
decisive test of the covariance-audit conclusion. The torque
diagnostic of Section~\ref{sec:torque} is a description of four
published points under a stated, simplified uncertainty treatment. It
identifies a discriminating experimental signature but does not, by
itself, establish the physical origin of the offset. The torque data
originate from a single apparatus and calibration chain, and a full
covariance matrix for the four differences has not been published.
The independent-Type~A treatment used here is therefore a 
simplification.

\paragraph{Covariance-scaling convention.}
The scaled statistic $\Qlam = \Qzero / \lhatMLE^{2}$ applies the
historical proportional inflation uniformly to the predictive
covariance of Equation~\eqref{eq:V0}, including the NIST internal
covariance $\mathbf{S}_{\mathrm{N}}$ and the historical-mean
uncertainty. This is a declared modelling convention under the view
that the excess dispersion characterising the historical archive is a
general property of how reported covariances in $G$ metrology
underestimate the true uncertainty of a measurement, and therefore
applies to a new $G$ measurement regardless of laboratory or
apparatus. The view is supported empirically, though not uniquely
established, by the internal NIST dispersion reported in
Section~\ref{sec:nist-check}: the four NIST determinations are
themselves internally discrepant relative to their reported covariance
at approximately the historical scale,
$\sqrt{Q_{\mathrm{int}}/\nu} = 2.89$ versus the historical
$\lhatMLE = 2.967$. An alternative, more conservative convention would
scale only the historical-mean uncertainty $\operatorname{Var}(\mhat)$
and leave $\mathbf{S}_{\mathrm{N}}$ unmodified, on the grounds that
the NIST systematic-error analysis may already account for effects
that historical measurements underestimated. The two conventions
answer different questions: the uniform convention asks whether the
later measurement is consistent with the historical archive treated as
a representative draw from a common systematic-error distribution,
whereas the mean-only convention asks whether the central value of
the later measurement is consistent with the historical archive
without granting that its internal covariance is similarly
underestimated. The present analysis adopts the uniform convention as
a declared working choice without claiming that the data uniquely
establish the choice, and the mean-only convention is retained
as a sensitivity in the reproducibility package.

% ============================================================
\section{Conclusions}
\label{sec:conclusions}

The quoted covariance scale for the CODATA-2018 archive of
gravitational-constant determinations is inadequate under the
proportional-scale model, requiring an inflation factor of
$\lhatMLE = 2.967$ even after the exclusion of the BIPM-14 entry
associated with the direct apparatus predecessor of the 2026 NIST
redetermination. Under this historically inferred model and the
declared uniform covariance-scaling convention, the correlated
four-configuration NIST result---highly improbable under the archive's
reported, unscaled covariance
($\Qzero = 42.625$, $p = 1.24 \times 10^{-8}$)---becomes statistically
unremarkable ($\Qlam = 4.840$, finite-sample $F$ $p = 0.382$;
plug-in $\chi^{2}_{4}$ $p = 0.304$). The free-minus-servo torque
differences published alongside the NIST redetermination favour a
common absolute torque offset across the four tested configurations of
$\qty{1.918 \pm 0.215}{\pico\newton\metre}$ over an equally simple,
torque-proportional alternative. The NIST and corrected BIPM Mark-II
contrasts have the same sign and comparable order of magnitude, but
the BIPM comparison is neither configuration-resolved nor an
independent test of the NIST torque model. A dual-mode torsion-balance
experiment in which the gravitational source-mass separation is
systematically varied would provide a direct experimental
discriminator between the two statistical scaling models considered
here.

% ============================================================
\section*{Code and data availability}

The complete numerical analysis, including all input data, covariance
matrices, and generated results reported in this paper, is provided in
a versioned reproducibility package with input-data integrity
verification and a documented correspondence between each reported
numerical result and its generating computation. 

% ============================================================

\end{document}